\documentclass[12pt,preprint]{aastex}

\slugcomment{Submitted to ApJ: 2003--08--11; accepted: }
\shorttitle{CO Measurements with $Copernicus$}
\shortauthors{Crenny \& Federman}

\received{2003 August 14}
\begin{document}
\title{Reanalysis of $Copernicus$ Measurements on Interstellar Carbon Monoxide}
\author{T. Crenny\altaffilmark{1}$^,$\altaffilmark{2}
and S.R. Federman\altaffilmark{1}}
\altaffiltext{1}{Department of Physics and Astronomy, University of Toledo, 
Toledo, OH 43606.}
\altaffiltext{2}{Department of Physics, Wheeling Jesuit University, 
Wheeling, WV 26003.}

\begin{abstract}

We used archival data acquired with the $Copernicus$ satellite to 
reexamine CO column densities because self-consistent oscillator 
strengths are now available.  Our focus is on lines of sight 
containing modest amounts of molecular species.  Our 
resulting column densities are small enough that self-shielding from 
photodissociation is not occurring in the clouds probed by the 
observations.  While our sample shows that the 
column densities of CO and H$_2$ are related, no correspondence with 
the CH column density is evident.  The case for the CH$^+$ column 
density is less clear.  Recent 
chemical models for these sight lines suggest that CH is mainly a 
by-product of CH$^+$ synthesis in low density gas.  The models are 
most successful in reproducing the amounts of CO in the densest 
sight lines.  Thus, much of the CO absorption must 
arise from denser clumps along the line of sight to account for the 
trend with H$_2$.

\end{abstract}

%\keywords{ISM: abundances --- ISM: molecules --- ultraviolet: ISM}

\section{Introduction}

Because CO is the second most abundant molecule in interstellar space, it 
plays a central role in our understanding of this environment.  Our focus 
is on diffuse molecular clouds, where the CO abundance is derived from 
absorption seen against the ultraviolet (UV) continuum of a background star.  
Comparisons between observational results and theoretical photochemical 
models provide the means to extract the physical conditions for the 
absorbing material.  In particular, gas densities, kinetic temperatures, 
and the flux of UV radiation penetrating into the cloud can be inferred.  
In this paper, we present a refined analysis of CO spectra obtained with 
the $Copernicus$ satellite, emphasizing sight lines with modest CO column 
densities ($\approx$ 10$^{13}$ cm$^{-2}$).  The data are especially useful 
for interpreting the CO abundance in regions where CH$^+$ chemistry 
leads to the production of other molecules 
(e.g., Draine \& Katz 1986; Zsarg\'{o} \& Federman 
2003) and before the effects of CO self shielding become important 
(see van Dishoeck \& Black 1988).

Recent developments prompted us to revisit these data, which were 
discussed previously by Jenkins et al. (1973), Morton \& Hu (1975), 
Federman et al. (1980), Snow \& Jenkins (1980), Allen, Snow, \& Jenkins 
(1990), and Federman et al. (1994).  Foremost, a self-consistent set of 
oscillator strengths ($f$-values) for the $A-X$ bands seen at wavelengths 
greater than 1200 \AA\ and Rydberg transitions involving 
the $B-X$ (0-0), $C-X$ (0-0), and $E-X$ (0-0) bands below 1200 \AA\ 
is emerging (e.g., Chan, Cooper, \& Brion 1993; Eidelsberg et al. 1999; 
Federman et al. 2001).  In other words, the same CO column density 
(or abundance) is derived regardless of the band(s) under study.  In the 
present work, the Rydberg transitions are studied.  Second, more 
sophisticated packages, such as NOAO's IRAF and profile fitting 
routines, provide us with the ability to 
reduce and analyze the data in a more precise way.  

Through these two developments, a more robust 
set of observational results arises.  The next 
Section describes the methods employed in treating the archival data.  This 
is followed by a general discussion of the observational results and by an 
analysis of relationships among line of sight column densities for comparison 
with our earlier conclusions (Federman et al. 1980; Federman et al. 1994).  
While it is not our aim to present a detailed chemical picture here, 
we note that Zsarg\'{o} \& Federman (2003) incorporated the present 
results in chemical analyses of molecule formation during CH$^+$ synthesis.

\section{Observations}

\subsection{$Copernicus$ Data}

We extracted high-resolution spectra taken with the U1 photomultiplier tube 
from the $Copernicus$ archive at the Multiwavelength Archive at the Space 
Telescope Science Institute.  The nominal spectral resolution was 0.05 
\AA.  The spectra covered absorption from the $B-X$ (0-0) band at 1150 
\AA, the $C-X$ (0-0) band at 1088 \AA, and the $E-X$ (0-0) band at 1076 \AA.  
Individual scans were examined; those completely encompassing the 
CO band and free from peculiarities were 
rebinned and summed to yield a final spectrum.  
The stellar continua were fitted with a low-order polynomial 
within the IRAF environment to produce rectified spectra for further 
analysis.  Examples of rectified spectra are shown in Figures 1 to 3.  
There are two points to note about these spectra.  First, the available 
spectral range was limited by the coverage of the spectral scan.  This is 
evident in the $o$ And spectrum near the $E-X$ band (see Fig. 3).  
Second, a chlorine line at 1088 \AA\ is the feature at longer 
wavelengths in Fig. 2.

Equivalent widths ($W_{\lambda}$) were measured off these spectra for 
comparison with earlier work and with our results of fitting the CO band 
profiles.  The values of $W_{\lambda}$ appear in Table 1; upper limits 
represent 3-$\sigma$ values.  The quoted uncertainties were determined 
from the root-mean-square (rms) deviations in the stellar continuum and 
the width of the feature at the 50\% level.  For upper limits, the 
width was taken to be 0.05 \AA\ or the apparent value for a 2-$\sigma$ 
`feature', whichever was larger.

There are three potential sources of background in coadded $Copernicus$ 
spectra, energetic particle counts, stray light, and scattered light.  
All archived spectra were corrected for particle background.  Procedures 
to eliminate stray light were introduced in data acquired after 19 April, 
1973.  For measurements before this date, we estimated the contribution 
from stray light by examining scans of saturated H$_2$ and 
N~{\small II} lines at nearby wavelengths.  In the worst cases (HD~21278, 
67 Oph, and $o$ And), about 30\% of the continuum level could be from 
stray light.  Since accidental blockage of the stray light source could 
have taken place in the CO scans, we did not apply any corrections, but 
instead note the three sight lines by different symbols in the plots 
described below.  Neither did we correct for scattered light, which 
could represent 5 to 10\% of the local continuum.

As note above, many of these spectra were analyzed by others.  For 
most bands, the earlier measurements of $W_{\lambda}$ 
(Morton \& Hu 1975; Federman et al. 1980)
and ours show relatively good agreement, consistent at the 2-$\sigma$ level.  
Essentially all are consistent at the quoted 3-$\sigma$ level.  The 
results for $\sigma$ Sco may agree as well, but we note that Allen et al. 
(1990) did not provide an estimate for the uncertainty in their 
measurement.  The main difference is that our results 
generally have smaller uncertainties because (1) all available data were 
used to produce final spectra and (2) the rms deviations in 
the stellar continua were more easy to quantify upon adoption of 
low-order polynomials for their shapes.  Our values for $W_{\lambda}$ 
tend to be smaller than earlier measures; this may also be due to our 
ability for improved fits to stellar continua.  Since Snow \& Jenkins 
(1980) only presented column densities obtained from fitting the CO 
band profiles, we compare their results with 
ours after describing our method for profile synthesis.

Each band was synthesized separately with the least-squares fitting 
routine (see Lambert et al. 1994) used in our laboratory study of 
Rydberg transitions (Federman et al. 2001).  The wavelengths for 
individual rotational lines were taken from the compilation of 
Eidelsberg et al. (1991), and we used the $f$-values of Federman 
et al. (2001) for the $B-X$ (0-0), $C-X$ (0-0), and $E-X$ (0-0) bands 
seen in our spectra.  Moreover, a $b$-value of 1 km s$^{-1}$, based 
on ultra-high-resolution studies ($\lambda$/$\delta\lambda$ $=$ 
500,000 to 900,000) of CH absorption in diffuse clouds (Crane, 
Lambert, \& Sheffer 1995; Crawford 1995), and an 
instrumental width of 0.05 \AA\ were adopted.  [A similar value of 
1.2 km s$^{-1}$ was adopted by Snow \& Jenkins (1980) in their 
study of sight lines in Scorpius.]  For each spectrum, 
the wavelength offset for the R(0) line and column density were varied 
until the difference between measured and synthesized spectra was 
minimized.  (The spectra shown in Fig. 1-3 have the R(0) line appearing 
at its laboratory wavelength.)  As indicated by the values of 
$W_{\lambda}$ in Table 1, the absorption is quite weak in most instances.  
Only in the cases of stronger absorption can one discern partially 
resolved rotational structure.  For spectra revealing stronger 
absorption (e.g., see Fig. 2), we were able to infer the rotational 
excitation temperature ($T_{ex}$) as well.  This was done in an iterative 
manner; for sight lines revealing absorption from more than one band, 
comparison of fitting results among bands was performed.  Otherwise, 
$T_{ex}$ was set at 4.0 K.  For many of the directions, more than one 
neutral gas component is seen in C~{\small I} spectra taken with 
the Goddard High Resolution Spectrograph (GHRS) on the 
{\it Hubble Space Telescope} (Zsarg\'{o} \& Federman 2003).  
Since these directions tend to have rather weak CO absorption, the inferred 
CO column densities are not especially sensitive to optical depth 
effects, regardless of the number of components or the adopted $b$-value.  
For example, a column density of $3 \times 10^{13}$ cm$^{-2}$, larger 
than found for most of the present sample, yields an optical depth at 
line center of $\approx$ 3 for R(0) in the strong $C-X$ band.

The results of our fits appear in Table 2.  The CO column density 
toward $\epsilon$ Per differs by a small amount ($\le$ 2\%) 
compared to that analyzed by Zsarg\'{o} \& Federman (2003) and so 
their conclusions are not altered.  The listed uncertainties for 
individual determinations of column density are based 
on the uncertainties in $W_{\lambda}$.
In cases where fits to more than one band were possible, the final 
column density was derived by taking a weighted mean of the individual 
determinations.  Fits to spectra showing no clear absorption in one 
band yielded consistent column densities.  However, the column 
density toward $\lambda$ Ori from the $W_{\lambda}$ for 
the $B-X$ band clearly is much larger than with those 
from the other bands and is not listed in Table 2.  In a 
similar vein, the relatively large upper limits determined for the $B-X$ 
band toward 1 Sco, $\nu$ Sco, $\mu$ Nor, and $\gamma$ Ara are not 
especially useful.  Finally, for sight lines with only upper 
limits, the most stringent column density was used in our analyses 
below.

We now compare the results of the fits with our direct measurements 
for $W_{\lambda}$ and with the column densities given by Snow \& Jenkins 
(1980) and Federman et al. (1994).  For the most 
part, the values of $W_{\lambda}$ derived from 
profile synthesis agree very well with our new measurements.  The largest 
differences in $W_{\lambda}$, usually at the 2-$\sigma$ level, occur 
for the $C-X$ band, such as that seen in Fig. 2 for 20 Tau.  
The band is partially blended with absorption from Cl~{\small I} 
at 1088 \AA, making these spectra the most susceptible to a poorly defined 
continuum.  The comparison with the column densities of Snow 
\& Jenkins is less satisfactory.  Their column densities are several 
times larger than ours.  The difference does not lie in the adopted 
$b$-value or $T_{ex}$; they are very similar.  Moreover, the $f$-values 
used in the fits, those quoted by Snow (1975) versus those of 
Federman et al. (2001), are not that different.  Other possible causes 
cannot be discerned because Snow \& Jenkins did not tabulate 
$W_{\lambda}$.  Federman et al. (1994) used published values for 
$W_{\lambda}$ and updated $f$-values for the $C-X$ and $E-X$ bands 
in their compilation.  Differences between their adopted values of 
$W_{\lambda}$ and our fitted ones, when combined with the factor of 1.6 
for the ratio of adopted $f$-values [those of Federman et al. (2001) being 
larger], successfully explains the differences seen in the present 
column densities and the ones quoted in the compilation.  In passing 
we note that Federman et al. (1994) also found self-consistent results 
from the two CO bands because the new and older $f$-values differ by a 
scale factor (1.6).

\subsection{Data from the {\it Hubble Space Telescope}}

In the course of our study on nonthermal chemistry in gas with low 
molecular abundances (Zsarg\'{o} \& Federman 2003), we found archival 
spectra for HD~112244 covering the $A-X$ (4-0) and (5-0) bands.  The 
data were acquired with grating G160M of the GHRS.  
The spectra (z0yv0607m, z0yv0608m, and z0yv0609m) 
were reduced in the manner described by Zsarg\'{o} \& Federman.  In 
addition, we smoothed the final, rectified spectra by three pixels to 
increase the signal to noise.  The bands were synthesized with the 
$f$-values of Chan et al. (1993), a $b$-value of 1 km s$^{-1}$, and 
$T_{ex}$ of 4.0 K.  The values for $W_{\lambda}$ and the column 
densities appear in Table 3.  Two components are discerned, though the 
weaker, bluer one is formally a 2.5-$\sigma$ detection.  Its reality 
is strengthened by the fact that the velocity separation of 
$\sim$ 25 km s$^{-1}$ agrees nicely with 
those seen in CH (Danks, Federman, \& Lambert 
1984) and C~{\small I} (Zsarg\'{o} \& Federman 2003), 24.6 and 24.1 
km s$^{-1}$, respectively.  As found here for CO, the red CH component 
is the stronger one, while the two components have similar strengths 
in C~{\small I}.

\section{Discussion}

\subsection{General Comments}

The results described in the last Section represent the lowest CO 
column densities measured in interstellar space.  Most determinations 
are in the range of $10^{12}$ to $10^{13}$ cm$^{-2}$.  Studies of 
CO emission at millimeter wavelengths generally are sensitive to 
column densities greater than $10^{15}$ cm$^{-2}$, values seen in 
molecule-rich diffuse gas like that toward $\zeta$ Oph (e.g., Lambert 
et al. 1994).  Even when measuring absorption at millimeter 
wavelengths against background extragalactic sources (e.g., Liszt \& 
Lucas 1998), column densities are found to be greater than $10^{14}$ 
cm$^{-2}$.  Clearly, UV absorption is the most sensitive probe of 
CO in diffuse clouds.  

This conclusion is consistent with another result arising from our 
profile syntheses.  We find that the CO excitation temperature is 
barely above the 2.7 K value of the Cosmic Background.  For 
comparison, $T_{ex}$ is about 4 to 6 K in molecule-rich diffuse gas 
(e.g., Lambert et al. 1994; Federman et al. 2003).  The gas probed 
by our observations reveal severe subthermal CO excitation, another 
indication that we are sampling a relatively low density environment.  
Analysis of C~{\small I} excitation (Zsarg\'{o} \& Federman 2003) 
for many of the same directions leads to density estimates of 10 to 
200 cm$^{-3}$, with the larger estimates associated with 
sight lines (in Scorpius) having $T_{ex}$ of 4 K.  
For most of our other sight lines, Jenkins, Jura, \& Loewenstein 
(1983) extracted pressures from C~{\small I} absorption seen in 
$Copernicus$ spectra.  When combined with the kinetic temperatures 
deduced from the $J$ $=$ 0 and 1 levels of H$_2$ (Savage et al. 1977), 
low densities are again inferred.

\subsection{Correlations among Species}

The correspondences between $N$(CO) with $N$(H$_2$), $N$(CH), and 
$N$(CH$^+$) for our set of directions appear in Figures 4-6.  The plots 
distinguish between components A and B toward HD~112244.  
The upper limit for $N$(CH$^+$) in component B is obtained by multiplying 
the uncertainty given by Lambert \& Danks (1986) for component A by 3.  
The H$_2$ column densities come from Savage et al. (1977).  The 
two components toward HD~112244 were given half the total H$_2$, 
consistent with the similar amounts of C~{\small I} in each 
(Zsarg\'{o} \& Federman 2003).  The results 
for CH come from the compilation of Federman et al. (1994), 
updated to include the column densities found by Crane et al. 
(1995), and those for CH$^+$ are from Federman (1982), 
Lambert \& Danks (1986), Crane et al. (1995), and 
Price et al. (2001).  The column densities used in this analysis 
appear in Table 4.  Typical uncertainities are 30\% for $N$(H$_2$) and 
10 to 30\% for the column densities of carbon-bearing molecules.  We 
adopted 3-$\sigma$ upper limits for non-detections.

The relationship involving CO and H$_2$ appears 
to show a reasonable correlation.  If we assume the upper limits are 
detections, the correlation coefficient ($r^2$) is 0.53.  
Since our set of data includes upper limits on CO, 
we also performed a linear regression with censored 
data, relying on the package ASURV-Rev. 1.1 (see Isobe, Feigelson, \& 
Nelson 1986; Isobe \& Feigelson 1990; La Valley, Isobe, \& Feigelson 
1992).  The Buckley-Jones method was employed because limits existed only for 
the dependent variable, log~$N$(CO).  The fit is shown in Fig. 4 as the 
solid line; the slope is $1.58 \pm 0.34$ and the intercept is $-$18.35.  The 
fitted results are nearly identical to those from a simple linear 
least-squares fit ($1.46 \pm 0.31$ and $-$16.0) because few data are 
represented by limits.  The slope is similar to the one found previously by 
Federman et al. (1980), whose survey included both molecule-rich and 
molecule-poor lines of sight.

The value of the slope arises from the competition between CO 
production under equilibrium conditions in colder, denser gas 
($n$ $\gtrsim 100$ cm$^{-3}$) and production as a result of CH$^+$ 
synthesis in lower density gas ($n$ $\lesssim 100$ cm$^{-3}$).  
Under equilibrium conditions in diffuse molecular clouds, CO is a 
second generation molecule, forming in appreciable quantities once 
significant amounts of OH are present (e.g., Federman \& Huntress 
1989).  In particular, the synthesis of CO mainly 
arises from reactions between C$^+$ and OH.  
If this were the only pathway leading to CO, a slope of 2 
would be expected (Federman et al. 1984).  On the other hand, when 
CH$^+$ is synthesized in lower density material in the cloud
via nonequilibrium processes, observable amounts 
of CO are produced in subsequent reactions involving CH$^+$ and O 
(e.g., Zsarg\'{o} \& Federman 2003).  Then $N$(CO) is not 
dependent on $N$(H$_2$).  The combination of the two schemes leads 
to a slope smaller than 2.  Some of the dispersion seen in Fig. 4 
probably results from differences in the relative contributions 
from the two pathways for CO.  For directions in common with the 
present study, Zsarg\'{o} \& Federman typically find 10 to 30\% of 
the CO comes from CH$^+$, except for the sight lines in Scorpius 
where most of the CO is attributed to the presence of 
CH$^+$ because the densities are several times higher.  Moreover, the 
modest relationship between CO and H$_2$ likely arises because 
most of the H$_2$ along the line of sight is associated with the denser 
material, not the gas containing CH$^+$.

Since there is a strong correlation between log~$N$(CH) and log~$N$(H$_2$) 
in diffuse molecular gas (Federman 1982; Danks et al. 1984), it is 
surprising at first glance to see such a poor correspondence between 
$N$(CO) and $N$(CH) (Fig. 5), where $r^2$ $=$ 0.21.  The correlation 
coefficient of 0.54 for log~$N$(CH) versus log~$N$(H$_2$) for our 
dataset indicates a tighter correspondence, but even this is significantly 
less than $r^2$ $=$ 0.80 found by Danks et al.  While a linear 
regression by the Schmitt Method in ASURV (which treats limits in the 
independent variable) suggests that 4 of 5 upper limits are consistent 
with detections, the slope and intercept are not 
well defined.  The relationship also contrasts with 
the one revealed in Fig. 9 of Federman et al. (1994).  
The key to understanding the present result lies 
in the source for CH in our sample.  
The sight lines in the current survey are not 
particularly rich in molecules in large part because the gas densities 
are rather low (e.g., Zsarg\'{o} \& Federman 2003).  Confirmation of 
low densities comes from observations of C$_2$ and CN, tracers of denser 
gas (Joseph et al. 1986; Federman et al. 1994), 
which yield only upper limits for the sight 
lines examined here.  Zsargo \& Federman found that 
most of the observed CH could be attributed 
to production of CH$^+$ under nonequilibrium conditions.  (Like CO, 
CH is produced under both equilibrium and nonequilibrium conditions.)  
Since reactions between CH and O play a minor role 
compared to those between CH$^+$ and O 
(Zsarg\'{o} \& Federman 2003), little correspondence between CO 
and CH is expected for our sight lines.  On the 
other hand, the sample in Federman et al. (1994) primarily contains 
molecule-rich clouds where CH chemistry is intimately tied to the 
amount of H$_2$ (Federman 1982; Danks et al. 1984).

The data in Fig. 6 suggest a correlation between CO and CH$^+$ as well 
($r^2$ $=$ 0.61 assuming all data are detections).  The slope is 
$1.31 \pm 0.24$, yet a linear relationship is expected when CO arises 
from CH$^+$ $+$ O (Federman et al. 1984).  The Schmitt Method indicates 
that only 2 of 8 upper limits are possible detections and yields a 
slope that is not well determined.  Therefore, it seems 
that the apparent correlation is strongly influenced by the results 
for $\pi$ Sco in the lower left corner of the plot.  Most of the CO 
appears to exist in regions denser than those responsible for the 
CH$^+$ (and CH) toward many stars in our sample.

In summary, we presented the most complete, internally consistent set 
of CO column densities for sight lines containing modest amounts of 
molecular material.  These column densities are 
relatively small; CO self-shielding does not affect the photodissociation 
rate needed to model these sight lines.  Thus our results can be 
used to test chemical aspects of the models that do not rely 
on radiative transfer in lines.  The correspondence between CO and 
H$_2$ found in earlier studies is still present, but there is no 
apparent connection with CH for these directions.  This 
latter result indicates that much of the CO is probing the denser portions 
of our sample of diffuse clouds, those clouds where CH is mainly produced 
as a by-product of CH$^+$ synthesis.  The ambiguous relationship between 
CO and CH$^+$ also seems to reflect different chemical environments.

\acknowledgments

We thank Matt Fritts for his help on codes used for profile synthesis, 
David Knauth for his assistance with IDL routines, and David Lambert 
and the referee, Don York, for comments on earlier versions of the paper.  
We acknowledge the useful exchanges with Don York, Ed Jenkins, and Jim 
Lauroesch regarding background levels in $Copernicus$ spectra.  
We utilized the Copernicus archive available at the 
Multiwavelength Archive at the Space Telescope Science Institute.  
Additional observations made with the NASA/ESA {\it Hubble Space Telescope} 
were obtained from the data archive at STScI.  STScI is operated by the 
Association of Universities for Research in Astronomy, Inc. under 
NASA contract NAS5-26555.  T. Crenny 
participated in the Research Experience for Undergraduates at the 
University of Toledo under NSF-REU Grant PHY-0097367.  
The research presented here was also supported in part by NASA Long 
Term Space Astrophysics grant NAG5-4957 to the University of Toledo.

\clearpage
\begin{deluxetable}{llccccccccccc}
%\rotate
\tablecolumns{13}
\tablewidth{0pt}
\tabletypesize{\scriptsize}
\tablecaption{CO Measurements with the $Copernicus$ Satellite}
\startdata
\hline \hline\\
 & & \multicolumn{11}{c}{$W_{\lambda}$ (m\AA)} \\ \cline{3-13}
HD & Star & \multicolumn{3}{c}{$B-X$ (1150 \AA)} & & 
\multicolumn{3}{c}{$C-X$ (1088 \AA)} & & 
\multicolumn{3}{c}{$E-X$ (1076 \AA)} \\ \cline{3-5} \cline{7-9}
\cline{11-13}
 & & Measured & Fit & Other & & Measured & Fit & Other & 
& Measured & Fit & Other \\ \hline
21278 & & $\ldots$ & $\ldots$ & $\ldots$ & & 
8.6$\pm$2.3 & 11.1 & 21$\pm$3\ $^a$ & & 2.5$\pm$0.8 & 5.3 & 7$\pm$5\ $^a$ \\
23408 & 20 Tau & $\ldots$ & $\ldots$ & $\ldots$ & &  
31.1$\pm$3.6 &25.9 & 32$\pm$8\ $^a$ & & $\ldots$ & $\ldots$ & $\ldots$ \\
23630 & $\eta$ Tau & $\ldots$ & $\ldots$ & $\ldots$ & & 
$\le$ 4.1 & $\le$ 3.2 & 2.8$\pm$2.4\ $^a$ & & $\ldots$ & $\ldots$ & $\ldots$ \\
24760 & $\epsilon$ Per & $\ldots$ & $\ldots$ & $\ldots$ & & 
1.4$\pm$0.3 & 1.4 & 8$\pm$2\ $^a$ & & 0.6$\pm$0.2 & 0.7 & 1.3$\pm$0.7\ $^a$ \\
30614 & $\alpha$ Cam & 12.9$\pm$2.6 & 16.8 & $\ldots$ & & 
67.6$\pm$6.0 & 46.4 & 44\ $^b$ & & $\ldots$ & $\ldots$ & $\ldots$ \\
36861 & $\lambda$ Ori & 1.9$\pm$0.2 & $\ldots$ & $\ldots$ & & 
3.9$\pm$0.7 & 4.7 & 24\ $^b$ & & 4.6$\pm$0.8 & 4.6 & $\ldots$ \\
40111 & 139 Tau & $\ldots$ & $\ldots$ & $\ldots$ & & 
$\le$ 2.8 & $\le$ 3.1 & 4.0$\pm$2.8\ $^a$ & & $\ldots$ & $\ldots$ & $\ldots$ \\
141637 & 1 Sco & $\le$ 11.0 & $\ldots$ & $\ldots$ & & 
$\le$ 3.8 & $\le$ 3.8 & $\ldots$ & & $\ldots$ & $\ldots$ & $\ldots$ \\
143018 & $\pi$ Sco & $\ldots$ & $\ldots$ & $\ldots$ & & 
0.8$\pm$0.2 & 1.3 & 0.8$\pm$0.2\ $^a$ & & $\ldots$ & $\ldots$ & $\ldots$ \\ 
143275 & $\delta$ Sco & $\ldots$ & $\ldots$ & $\ldots$ & & 3.6$\pm$1.1 
& 4.5 & 6.0$\pm$0.8\ $^a$,\ $^c$ & & $\ldots$ & $\ldots$ & $\ldots$ \\
144217 & $\beta^1$ Sco & $\le$ 1.5 & $\le$ 0.8 & $\ldots$ & & 
5.4$\pm$0.6 & 6.0 & $^c$ & & $\ldots$ & $\ldots$ & $\ldots$ \\
144470 & $\omega^1$ Sco & 1.1$\pm$0.2 & 1.0 & $^c$ & & 8.5$\pm$0.9 
& 10.4 & 12$\pm$1.6\ $^a$,\ $^c$ & & $\ldots$ & $\ldots$ & $\ldots$ \\
145502 & $\nu$ Sco & $\le$ 6.1 & $\ldots$ & $\ldots$ & & 5.9$\pm$0.6 
& 7.5 & 5.0$\pm$2.2\ $^a$,\ $^c$ & & $\ldots$ & $\ldots$ & $\ldots$ \\
147165 & $\sigma$ Sco & $\ldots$ & $\ldots$ & $\ldots$ & & 
3.7$\pm$0.3 & 4.4 & 6\ $^d$ & & $\ldots$ & $\ldots$ & $\ldots$ \\
149038 & $\mu$ Nor & $\le$ 17.0 & $\ldots$ & $\ldots$ & & 
30.7$\pm$3.8 & 30.2 & 39$\pm$7\ $^a$ & & $\ldots$ & $\ldots$ & $\ldots$ \\
157246 & $\gamma$ Ara & $\le$ 3.6 & $\ldots$ & $\ldots$ & & 2.1$\pm$0.2 
& 2.6 & 3.2$\pm$0.8\ $^e$ & & 1.6$\pm$0.3 & 1.7 & 2.6$\pm$0.7\ $^e$ \\
164353 & 67 Oph & $\ldots$ & $\ldots$ & $\ldots$ & & 13.0$\pm$2.1 
& 12.9 & 10$\pm$7\ $^a$ & & 6.7$\pm$1.2 & 6.3 & 8$\pm$4\ $^a$ \\
200120 & 59 Cyg & $\ldots$ & $\ldots$ & $\ldots$ & & 1.5$\pm$0.4 & 
2.3 & 5.0$\pm$1.5\ $^a$ & & $\le$ 1.1 & $\le$ 0.9 & 0.9$\pm$1.4\ $^a$ \\
217675 & $o$ And & $\ldots$ & $\ldots$ & $\ldots$ & & 
5.1$\pm$0.7 & 6.4 & 12$\pm$2\ $^a$ & & 4.8$\pm$1.2 & 5.1 & 10$\pm$3\ $^a$ \\
218376 & 1 Cas & $\ldots$ & $\ldots$ & $\ldots$ & & 
58.7$\pm$12.0 & 52.7 & $\ldots$ & & $\ldots$ & $\ldots$ & $\ldots$ \\ \hline
\enddata
\tablenotetext{a}{Federman et al. 1980.}
\tablenotetext{b}{Jenkins et al. 1973 did not give $W_{\lambda}$, but 
assumed the absorption was on the linear portion of the curve of growth.}
\tablenotetext{c}{Snow \& Jenkins 1980 did not give $W_{\lambda}$.}
\tablenotetext{d}{Allen et al. 1990.}
\tablenotetext{e}{Morton \& Hu 1975.}
\end{deluxetable}

\begin{deluxetable}{lccccc}
%\rotate
\tablecolumns{6}
\tablewidth{0pt}
\tabletypesize{\scriptsize}
\tablecaption{Results of Profile Synthesis\ $^a$}
\startdata
\hline \hline\\
Star & \multicolumn{4}{c}{$N$(CO) (cm$^{-2}$)} & $T_{ex}$ (K) \\ \cline{2-5}
 & $B-X$ & $C-X$ & $E-X$ & Final & \\ \hline
HD~21278 & $\ldots$ & $(9.6\pm5.1) \times 10^{12}$ & 
$(7.0\pm2.2) \times 10^{12}$ & $(7.4\pm2.0) \times 10^{12}$ & 4.0 \\
20 Tau & $\ldots$ & $(5.4\pm0.6) \times 10^{13}$ & $\ldots$ & 
$(5.4\pm0.6) \times 10^{13}$ & 3.0 \\
$\eta$ Tau & $\ldots$ & $\le$ $2.2 \times 10^{12}$ & $\ldots$ & 
$\le$ $2.2 \times 10^{12}$ & 4.0 \\
$\epsilon$ Per & $\ldots$ & $(8.8\pm1.9) \times 10^{11}$ & 
$(8.1\pm2.7) \times 10^{11}$ & $(8.6\pm1.6) \times 10^{11}$ & 3.0 \\
$\alpha$ Cam & $(3.1\pm0.6) \times 10^{14}$ & $(3.1\pm0.3) \times 10^{14}$ & 
$\ldots$ & $(3.1\pm0.3) \times 10^{14}$ & 3.3 \\
$\lambda$ Ori & $^b$ & $(3.2\pm0.6) \times 10^{12}$ & 
$(5.8\pm1.0) \times 10^{12}$ & $(3.9\pm0.5) \times 10^{12}$ & 3.2 \\
139 Tau & $\ldots$ & $\le$ $2.1 \times 10^{12}$ & $\ldots$ & 
$\le$ $2.1 \times 10^{12}$ & 4.0 \\
1 Sco & $^b$ & $\le$ $2.6 \times 10^{12}$ & $\ldots$ & 
$\le$ $2.6 \times 10^{12}$ & 4.0 \\
$\pi$ Sco & $\ldots$ & $(8.3\pm2.0) \times 10^{11}$ & $\ldots$ & 
$(8.3\pm2.0) \times 10^{11}$ & 4.0 \\ 
$\delta$ Sco & $\ldots$ & $(3.1\pm0.9) \times 10^{12}$ & $\ldots$ & 
$(3.1\pm0.9) \times 10^{12}$ & 4.0 \\
$\beta^1$ Sco & $\le$ $8.1 \times 10^{12}$ & $(4.3\pm0.5) \times 10^{12}$ & 
$\ldots$ & $(4.3\pm0.5) \times 10^{12}$ & 4.0 \\
$\omega^1$ Sco & $(10.0\pm2.0) \times 10^{12}$ & $(8.8\pm1.0) \times 10^{12}$ & 
$\ldots$ & $(9.0\pm0.9) \times 10^{12}$ & 4.0 \\
$\nu$ Sco & $^b$ & $(5.7\pm0.6) \times 10^{12}$ & $\ldots$ & 
$(5.7\pm0.6) \times 10^{12}$ & 3.2 \\
$\sigma$ Sco & $\ldots$ & $(3.0\pm0.2) \times 10^{12}$ & $\ldots$ & 
$(3.0\pm0.2) \times 10^{12}$ & 4.0 \\
$\mu$ Nor & $^b$ & $(8.1\pm1.0) \times 10^{13}$ & $\ldots$ & 
$(8.1\pm1.0) \times 10^{13}$ & 3.0 \\
$\gamma$ Ara & $^b$ & $(1.7\pm0.2) \times 10^{12}$ & 
$(2.0\pm0.4) \times 10^{12}$ & $(1.8\pm0.2) \times 10^{12}$ & 3.2 \\
67 Oph & $\ldots$ & $(1.3\pm0.2) \times 10^{13}$ & 
$(0.8\pm0.3) \times 10^{13}$ & $(1.1\pm0.2) \times 10^{13}$ & 3.0 \\
59 Cyg & $\ldots$ & $(1.5\pm0.4) \times 10^{12}$ & 
$\le$ $1.1 \times 10^{12}$ & $(1.5\pm0.4) \times 10^{12}$ & 3.0 \\
$o$ And & $\ldots$ & $(4.7\pm1.6) \times 10^{12}$ & 
$(6.6\pm1.6) \times 10^{12}$ & $(5.6\pm1.1) \times 10^{12}$ & 3.5 \\
1 Cas & $\ldots$ & $(6.3\pm1.3) \times 10^{14}$ & $\ldots$ & 
$(6.3\pm1.3) \times 10^{14}$ & 3.0 \\ \hline
\enddata
\tablenotetext{a}{The adopted $f$-values for the $B-X$, $C-X$, and $E-X$ 
bands are $6.7 \times 10^{-3}$, $1.23 \times 10^{-1}$, and 
$6.8 \times 10^{-2}$ (Federman et al. 2001).}
\tablenotetext{b}{Fit does not provide useful constraint.}
\end{deluxetable}

\begin{deluxetable}{lccccccc}
%\rotate
\tablecolumns{8}
\tablewidth{0pt}
\tabletypesize{\scriptsize}
\tablecaption{Results for HD~112244}
\startdata
\hline \hline\\
Component & \multicolumn{3}{c}{$A-X$ (4-0)} & & 
\multicolumn{3}{c}{$A-X$ (5-0)} \\ \cline{2-4} \cline{6-8}
 & $W_{\lambda}$(obs) & $W_{\lambda}$(fit) & $N$(CO)\ $^a$ & & 
$W_{\lambda}$(obs) & $W_{\lambda}$(fit) & $N$(CO)\ $^a$ \\
 & (m\AA) & (m\AA) & (cm$^{-2}$) & & (m\AA) & (m\AA) & (cm$^{-2}$) \\ \hline
A\ $^b$ & 3.9$\pm$0.7 & 4.3 & $(1.1\pm0.2) \times 10^{13}$ & & 
1.9$\pm$0.6 & 2.2 & $(0.9\pm0.3) \times 10^{13}$ \\
B & 1.2$\pm$0.5 & 1.2 & $(3.0\pm1.3) \times 10^{12}$ & & 
1.6$\pm$0.6 & 1.6 & $(6.4\pm2.4) \times 10^{12}$ \\ \hline
\enddata
\tablenotetext{a}{The weighted final column densities for components 
A and B are $(1.0\pm0.2) \times 10^{13}$ and $(3.8\pm1.1) \times 10^{12}$ 
cm$^{-2}$, respectively.}
\tablenotetext{b}{Component A is the redder of the two.}
\end{deluxetable}

\begin{deluxetable}{lcccc}
%\rotate
\tablecolumns{5}
\tablewidth{0pt}
\tabletypesize{\scriptsize}
\tablecaption{Molecular Column Densities}
\startdata
\hline \hline\\
Star & $N$(CO) & $N$(H$_2$) & $N$(CH) & $N$(CH$^+$) \\
 & (cm$^{-2}$) & (cm$^{-2}$) & (cm$^{-2}$) & (cm$^{-2}$) \\ \hline
HD~21278 & $7.4 \times 10^{12}$ & $3.0 \times 10^{19}$ & 
$\ldots$ & $4.8 \times 10^{12}$ \\
20 Tau & $5.4 \times 10^{13}$ & $5.6 \times 10^{19}$ & 
$\le$ $3.6 \times 10^{11}$ & $2.6 \times 10^{13}$ \\
$\eta$ Tau & $\le$ $2.2 \times 10^{12}$ & $3.5 \times 10^{19}$ & 
$\ldots$ & $2.0 \times 10^{12}$ \\
$\epsilon$ Per & $8.6 \times 10^{11}$ & $3.3 \times 10^{19}$ & 
$\le$ $3.0 \times 10^{12}$ & $\le$ $1.6 \times 10^{12}$ \\
$\alpha$ Cam & $3.1 \times 10^{14}$ & $2.2 \times 10^{20}$ & 
$6.8 \times 10^{12}$ & $2.0 \times 10^{13}$ \\
$\lambda$ Ori & $3.9 \times 10^{12}$ & $1.3 \times 10^{19}$ & 
$\ldots$ & $\le$ $1.0 \times 10^{12}$ \\
139 Tau & $\le$ $2.1 \times 10^{12}$ & $5.5 \times 10^{19}$ & 
$\ldots$ & $\ldots$ \\
HD~112244A\ $^a$ & $1.0 \times 10^{13}$ & $7.5 \times 10^{19}$ & 
$3.0 \times 10^{12}$ & $7.5 \times 10^{12}$ \\ 
HD~112244B\ $^a$ & $3.8 \times 10^{12}$ & $7.5 \times 10^{19}$ & 
$2.1 \times 10^{12}$ & $\le$ $3.5 \times 10^{12}$ \\ 
1 Sco & $\le$ $2.6 \times 10^{12}$ & $1.7 \times 10^{19}$ & 
$\ldots$ & $\le$ $2.1 \times 10^{12}$ \\
$\pi$ Sco & $8.3 \times 10^{11}$ & $2.1 \times 10^{19}$ & 
$\le$ $7.5 \times 10^{11}$ & $5.7 \times 10^{11}$ \\ 
$\delta$ Sco & $3.1 \times 10^{12}$ & $2.6 \times 10^{19}$ & 
$2.2 \times 10^{12}$ & $3.3 \times 10^{12}$ \\
$\beta^1$ Sco & $4.3 \times 10^{12}$ & $6.8 \times 10^{19}$ & 
$2.1 \times 10^{12}$ & $5.4 \times 10^{12}$ \\
$\omega^1$ Sco & $9.0 \times 10^{12}$ & $1.1 \times 10^{20}$ & 
$3.2 \times 10^{12}$ & $6.2 \times 10^{12}$ \\
$\nu$ Sco & $5.7 \times 10^{12}$ & $7.8 \times 10^{19}$ & 
$5.9 \times 10^{12}$ & $6.3 \times 10^{12}$ \\
$\sigma$ Sco & $3.0 \times 10^{12}$ & $6.2 \times 10^{19}$ & 
$3.1 \times 10^{12}$ & $5.8 \times 10^{12}$ \\
$\mu$ Nor & $8.1 \times 10^{13}$ & $2.8 \times 10^{20}$ & 
$1.0 \times 10^{13}$ & $3.5 \times 10^{13}$ \\
$\gamma$ Ara & $1.8 \times 10^{12}$ & $1.7 \times 10^{19}$ & 
$\le$ $9.0 \times 10^{11}$ & $\le$ $1.7 \times 10^{12}$ \\
67 Oph & $1.1 \times 10^{13}$ & $1.8 \times 10^{20}$ & 
$4.5 \times 10^{12}$ & $7.4 \times 10^{12}$ \\
59 Cyg & $1.5 \times 10^{12}$ & $2.0 \times 10^{19}$ & 
$\ldots$ & $\le$ $5.3 \times 10^{12}$ \\
$o$ And & $5.6 \times 10^{12}$ & $4.7 \times 10^{19}$ & 
$\le$ $1.9 \times 10^{12}$ & $\le$ $1.4 \times 10^{12}$ \\
1 Cas & $6.3 \times 10^{14}$ & $1.4 \times 10^{20}$ & 
$7.6 \times 10^{12}$ & $1.1 \times 10^{13}$ \\ \hline
\enddata
\tablenotetext{a}{Since the C~I column densities are so similar, 
we divided the total $N$(H$_2$) in half for each component.}
\end{deluxetable}

\clearpage
\begin{figure}
\epsscale{0.67}
\plotone{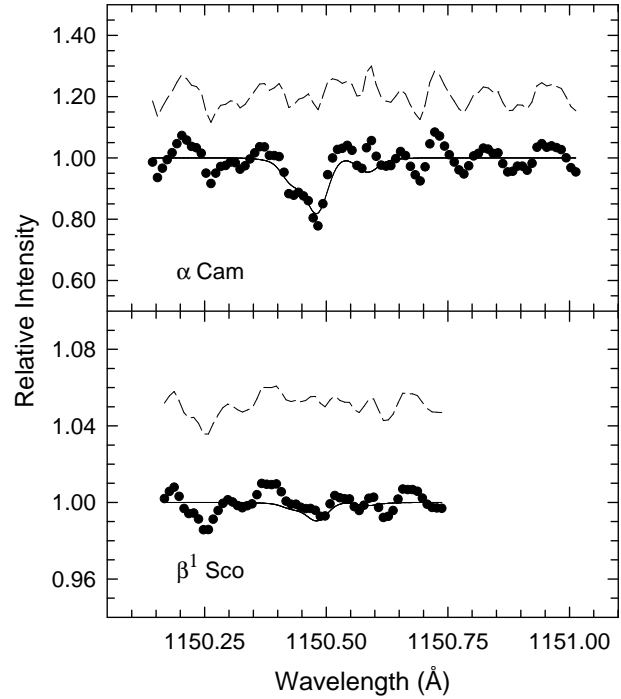}
\caption{Examples of fits to the $B-X$ bands in spectra for $\alpha$ 
Cam (upper panel) and $\beta^1$ Sco (lower panel).  The data are 
represented by filled circles, and the fit from the profile synthesis 
is the solid line through the data points.  The dashed line shows 
the difference between fit and data, offset to 1.20 and 1.05 in the 
spectrum for $\alpha$ Cam and $\beta^1$ Sco, respectively.  The spectrum 
for $\beta^1$ Sco, which is shown on an expanded vertical scale, 
yields an upper limit to the amount of CO absorption.}
\end{figure}

\begin{figure}
\epsscale{0.67}
\plotone{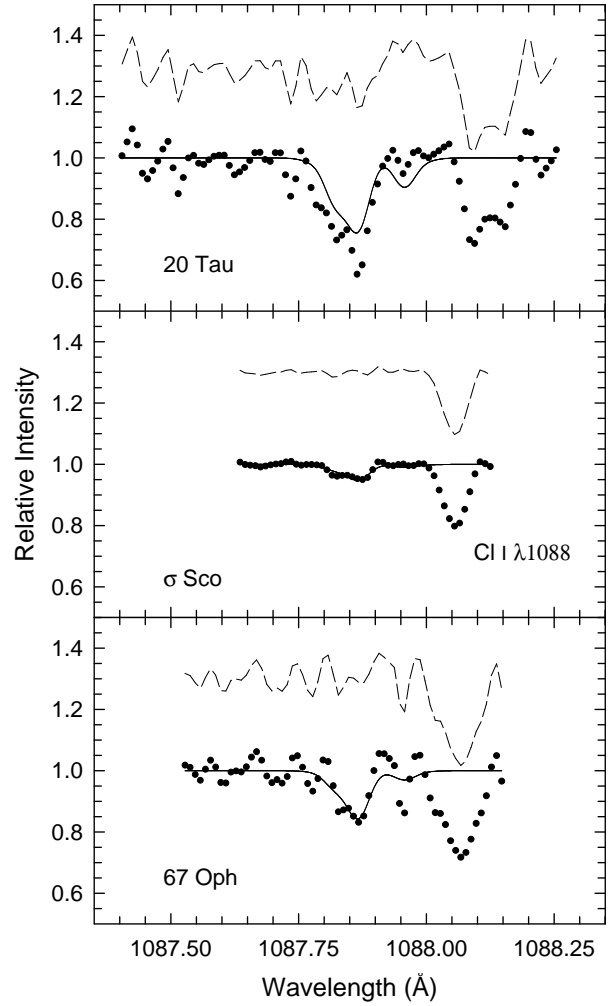}
\caption{Same as Fig. 1 for $C-X$ bands seen toward 20 Tau (upper panel), 
$\sigma$ Sco (middle panel), and 67 Oph (lower panel).  Here, the fit-data 
is offset to 1.30 in all panels.  The feature near 1088.05 \AA\ is a line 
from Cl~{\small I}.}
\end{figure}

\begin{figure}
\epsscale{0.67}
\plotone{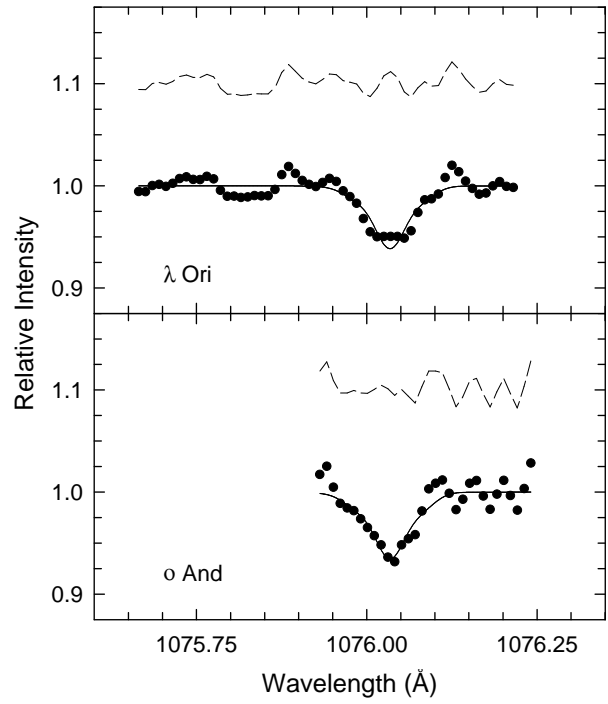}
\caption{Same as Fig. 1 for the $E-X$ bands at 1076 \AA\ seen toward 
$\lambda$ Ori (upper panel) and $o$ And (lower panel).  Here, the 
offset for fit-data is 1.10.}
\end{figure}

\begin{figure}
\epsscale{0.67}
\plotone{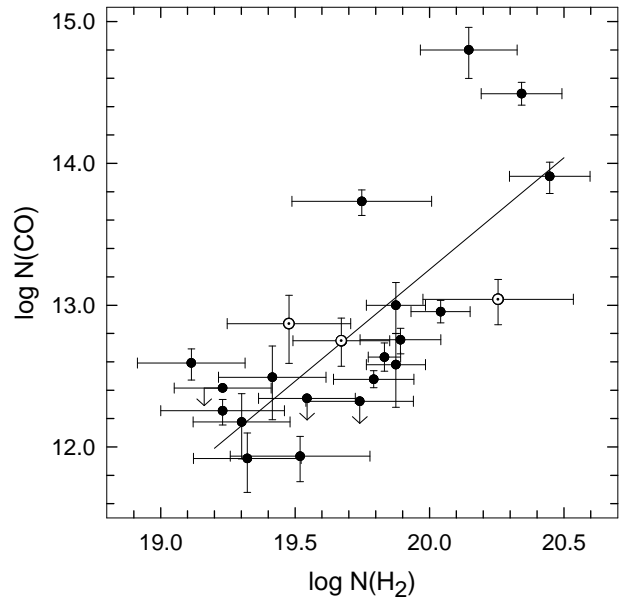}
\caption{Plot of log $N$(CO) vs. log $N$(H$_2$) for the sight lines in the 
present survey.  Results for HD~21278, 67 Oph, and $o$ And, where a 30\% 
(0.11 dex) contribution from stray light may be present, are indicated 
by dotted open circles.  The 2-$\sigma$ observational uncertainties are 
shown.  The solid line is the fit from the linear regression analysis with 
censored data $-$ i.e., upper limits.}
\end{figure}

\begin{figure}
\epsscale{0.67}
\plotone{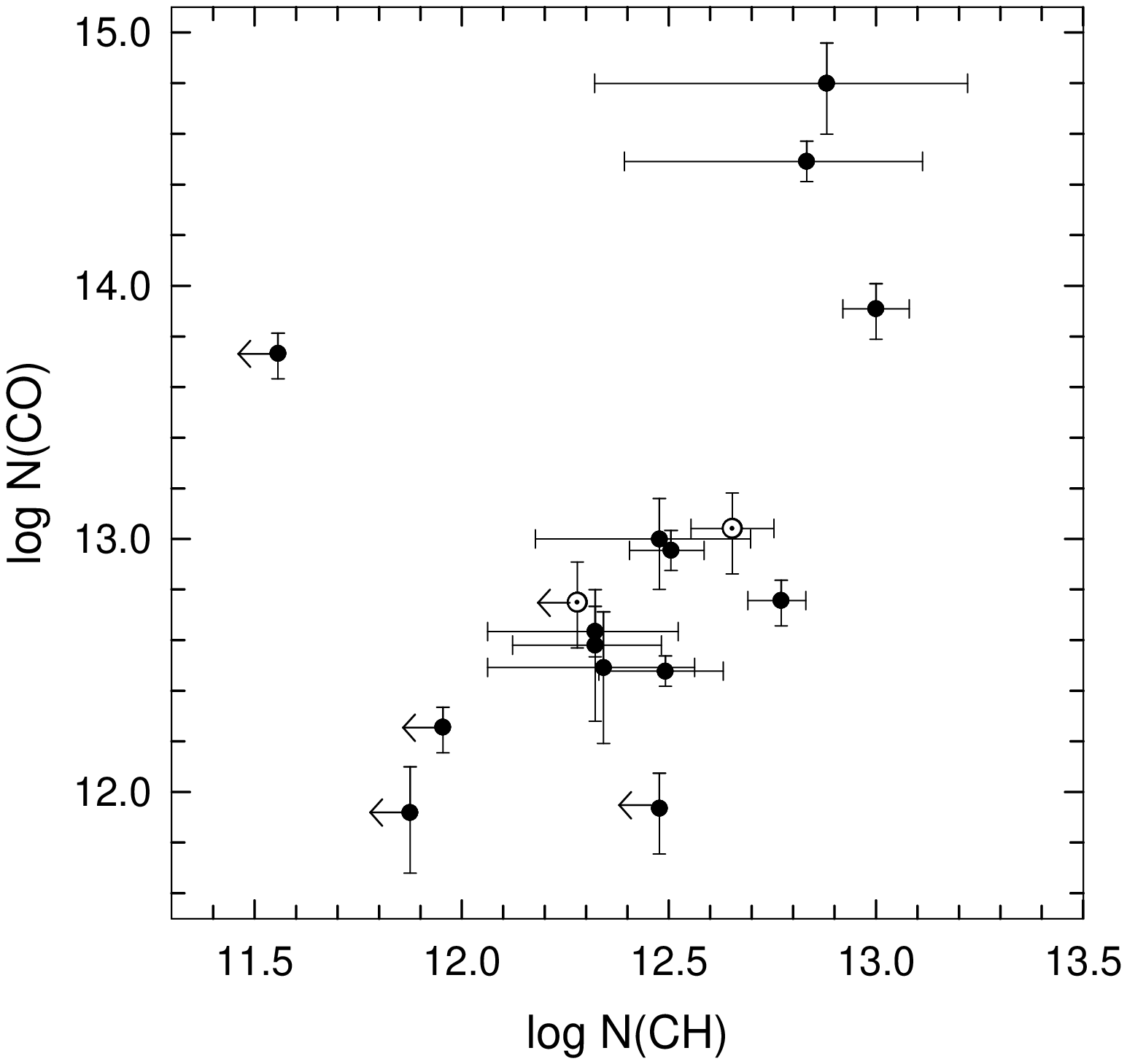}
\caption{Plot of log $N$(CO) vs. log $N$(CH) for the sight lines in the 
present survey.}
\end{figure}

\begin{figure}
\epsscale{0.67}
\plotone{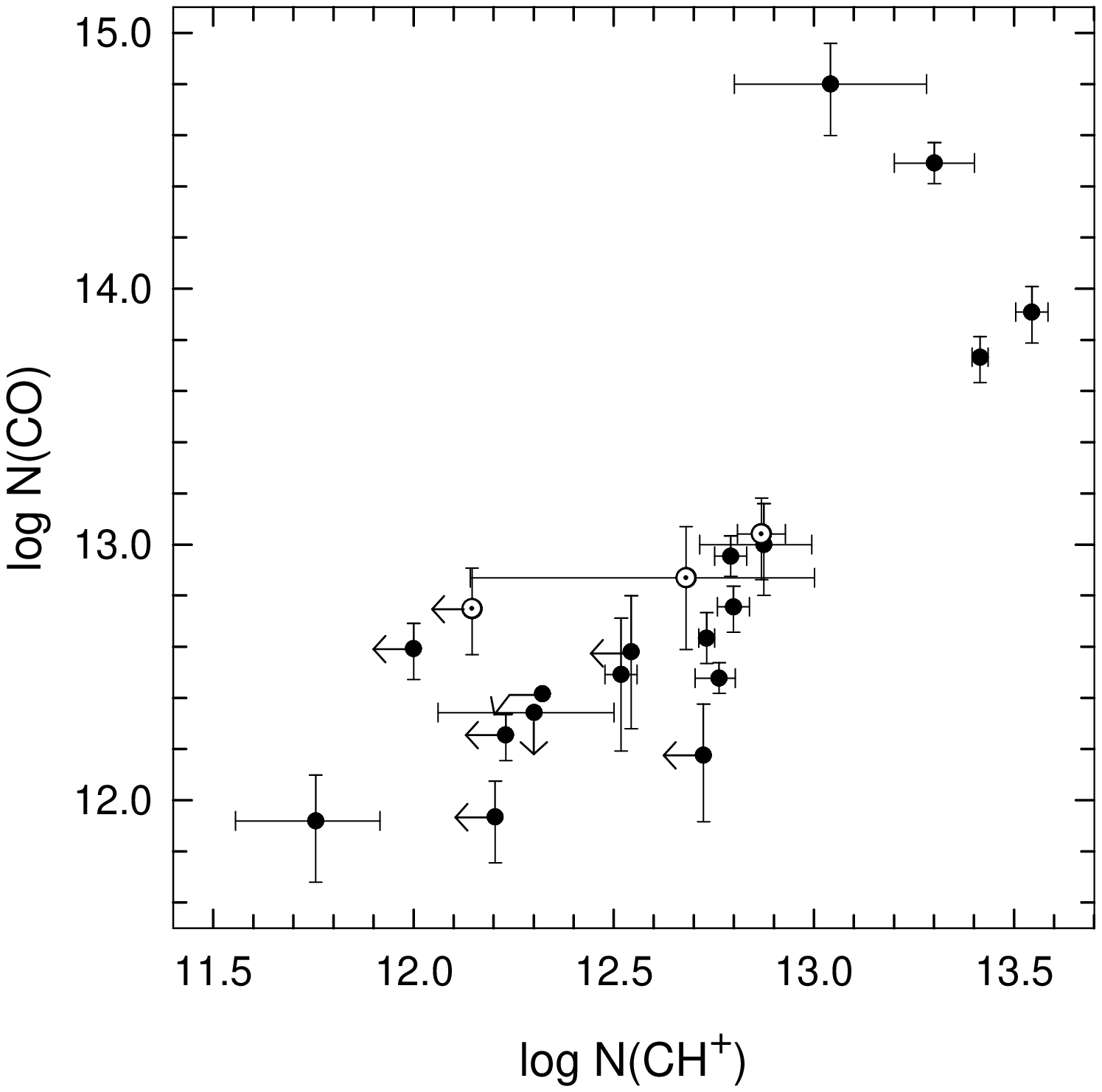}
\caption{Plot of log $N$(CO) vs. log $N$(CH$^+$) for the sight lines in the 
present survey.}
\end{figure}

\end{document}